\documentclass[superscriptaddress,aps,amsmath,amssymb,floatfix,preprint,nobalancelastpage,raggedbottom]{revtex4-1}
\usepackage{graphicx}
\usepackage[colorlinks=true,citecolor=blue,linkcolor=blue,urlcolor=blue]{hyperref}
\usepackage{amsmath}
\usepackage{dcolumn}
\usepackage{xcolor}
\usepackage{fancyhdr}
\usepackage{lipsum}
\usepackage{soul}

\usepackage{footnote}

\makeatletter 
\renewcommand{\fnum@figure}{\textbf{FIG.~\thefigure}}
\makeatother

\makeatletter
\def\bbordermatrix#1{\begingroup \m@th
  \@tempdima 4.75\p@
  \setbox\z@\vbox{%
    \def\cr{\crcr\noalign{\kern2\p@\global\let\cr\endline}}%
    \ialign{$##$\hfil\kern2\p@\kern\@tempdima&\thinspace\hfil$##$\hfil
      &&\quad\hfil$##$\hfil\crcr
      \omit\strut\hfil\crcr\noalign{\kern-\baselineskip}%
      #1\crcr\omit\strut\cr}}%
  \setbox\tw@\vbox{\unvcopy\z@\global\setbox\@ne\lastbox}%
  \setbox\tw@\hbox{\unhbox\@ne\unskip\global\setbox\@ne\lastbox}%
  \setbox\tw@\hbox{$\kern\wd\@ne\kern-\@tempdima\left[\kern-\wd\@ne
    \global\setbox\@ne\vbox{\box\@ne\kern2\p@}%
    \vcenter{\kern-\ht\@ne\unvbox\z@\kern-\baselineskip}\,\right]$}%
  \null\;\vbox{\kern\ht\@ne\box\tw@}\endgroup}
\makeatother

\begin{document}
\title{Ultrafast Spin-Transfer-Torque Switching of Synthetic Ferrimagnets}
\author{Kerem Yunus Camsari}
\email{kcamsari@purdue.edu}      
\affiliation{School of Electrical and Computer Engineering, Purdue University, IN, 47907}
\author{Ahmed Zeeshan Pervaiz}
\affiliation{School of Electrical and Computer Engineering, Purdue University, IN, 47907}
\author{Rafatul Faria}
\affiliation{School of Electrical and Computer Engineering, Purdue University, IN, 47907}
\author{Ernesto E. Marinero}
\email{eemarinero@purdue.edu}   
\affiliation{School of Electrical and Computer Engineering, Purdue University, IN, 47907}
\affiliation{School of Materials Engineering, Purdue University, IN, 47907}
\author{Supriyo Datta}
\email{datta@purdue.edu}      
\affiliation{School of Electrical and Computer Engineering, Purdue University, IN, 47907}
\date{\today}

\begin{abstract}
The switching speed and the write current required for spin-transfer-torque reversal of spintronic devices such as magnetic tunnel junctions (MTJ) currently hinder their wide implementation into memory and logic devices. This problem is further exacerbated as the dimensions of MTJ nanostructures are scaled down to tens of nanometers in diameter, as higher magnetic anisotropy materials are required to meet thermal stability requirements that demand higher switching current densities. Here, we  propose a simple solution to these  issues based on synthetic ferrimagnet (SFM) structures. It is commonly assumed that to achieve a given switching delay, the current has to exceed the critical current by a certain factor and so a higher critical current implies a higher switching current. We show that this is not the case for SFM structures which can provide significantly reduced switching delay for a given current density, even though the critical current is increased. This non-intuitive result can be understood from the requirements of angular momentum conservation. We conclude that a 20 nm diameter MTJ incorporating the proposed SFM free layer structure can be switched in tens of picosecond time scales. This remarkable switching speed can be attained employing current perpendicular magnetic anisotropy materials with experimentally demonstrated exchange coupling strengths.  
\end{abstract}
\maketitle

\section{Introduction}
\subsection{Exchange Coupled Magnets}
Exchange coupled magnets have been successfully used in the magnetic recording industry to reduce the magnetic field switching threshold of  high magnetic anisotropy materials by coupling them ferromagnetically to lower anisotropy materials \cite{victora2005composite,victora2005exchange}. These so-called, ``exchange coupled spring magnets'' or ``graded-anisotropy ferromagnets'' enable the reduction of the magnetic grain volume, a necessary requirement for ultra-high density recording, while retaining their thermal stability \cite{richter2007transition}. Exchange coupled structures are key constituents of commercially available ultra-high magnetic recording density hard disks. 

Inspired by the success of magnetic field-driven magnetic multilayers, there has been a growing interest in spin-transfer-torque driven synthetic structures. Synthetic ferro- and anti-ferromagnetically coupled magnetic layers have been studied as potential replacements of single ferromagnetic free layers in Magnetic Tunnel Junction (MTJ) stacks in experimental studies \cite{yakata2009thermal,hayakawa2008current}. Synthetic antiferromagnets have been of special interest due to the inherent advantages of antiferromagnets such as stray-field free magnetic stacks that can potentially operate at THz frequencies \cite{basset2008toward,macdonald2011antiferromagnetic,duine_2011}. Spin-torque nano oscillators based on synthetic antiferromagnets have been proposed theoretically \cite{klein2012interplay,zhou2013field}. In addition, spin-transfer-torque driven magnetic structures comprising high and low magnetic anisotropy materials have  been theoretically analyzed \cite{yulaev2011spin,victora2012exchange} and experimentally investigated    \cite{meng2005spin,meng2006composite,yen2008reduction}. These studies verified that the critical switching current of  ferromagnetically coupled magnetic bilayers can be reduced significantly. This is achieved by a judicious selection of the magnetic properties of the constituent layers and by controlling the interlayer exchange coupling strength by adjusting the thickness of spacers, such as Ruthenium, Rhodium and Ru-alloys or by intercalating magnetic alloys to achieve the desired exchange coupling strength. For maximum reduction of the critical current to switch these systems, the  exchange coupled layers are required to exhibit widely different magnetic properties: anisotropy, saturation magnetization and damping coefficients;  making their experimental implementation challenging if not prohibitive.

\subsection{Summary of this paper}
\begin{figure}[b]
\begin{center}
\includegraphics[width=\linewidth]{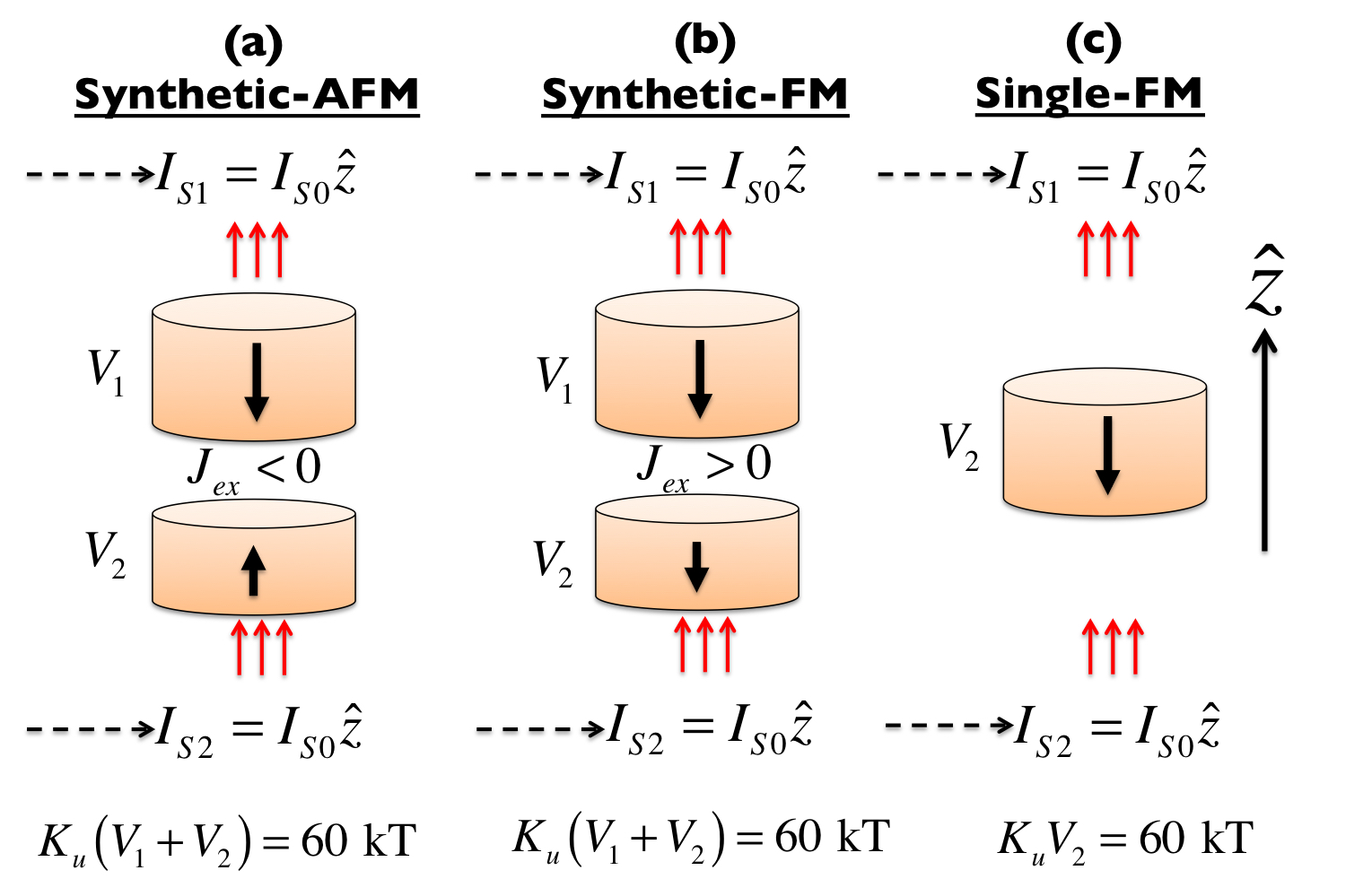}
\caption{\textbf{Magnetic Configurations:} We consider three structures that are driven by symmetrically polarized spin-currents: (a) A synthetic AFM layer comprised of two monodomain magnets having equal magnetic properties (anisotropy constant, magnetization per unit volume and damping coefficient, $K_u$, $M_s$, $\alpha$ respectively) but with different volumes, $V_1>V_2$ in our notation. (b) Synthetic FM layer driven by symmetrically polarized spin currents. (c) A single monodomain magnet driven by symmetrically spin-polarized currents. All magnet configurations are assumed to have equal thermal stability, for the exchange coupled magnetic configurations, we assume that the thermal stability is the sum of its constituent layers \cite{fullerton2000antiferromagnetically}.}
\label{fi:fig1}
\end{center}
\end{figure}

FIG.~(\ref{fi:fig1}) shows the three free layer structures that we analyze in this paper, our main focus being the synthetic ferrimagnets illustrated in FIG.~(\ref{fi:fig1}a). Synthetic ferromagnets and single ferromagnetic (FM) layers are shown for comparison in FIG.~(\ref{fi:fig1}b) and FIG.~(\ref{fi:fig1}c) respectively. Material parameters for the single FM are chosen to be comparable to those described in  \cite{kent2015new}. All the structures analyzed have the same total thermal stability ($\Delta= \rm 60 \ kT$) and we assume that this is equal to the sum of those of the constituent layers \cite{carey2001magnetic,fullerton2000antiferromagnetically}. The actual overall thermal stability of such coupled systems may involve additional considerations, but these are outside the scope of this paper \cite{taniguchi2012theoretical}. Thermal activation, noise and multi-domain effects are not considered in our treatment \cite{gopman2014switching}.

\begin{figure}[b]
\begin{center}
\includegraphics[width=5in]{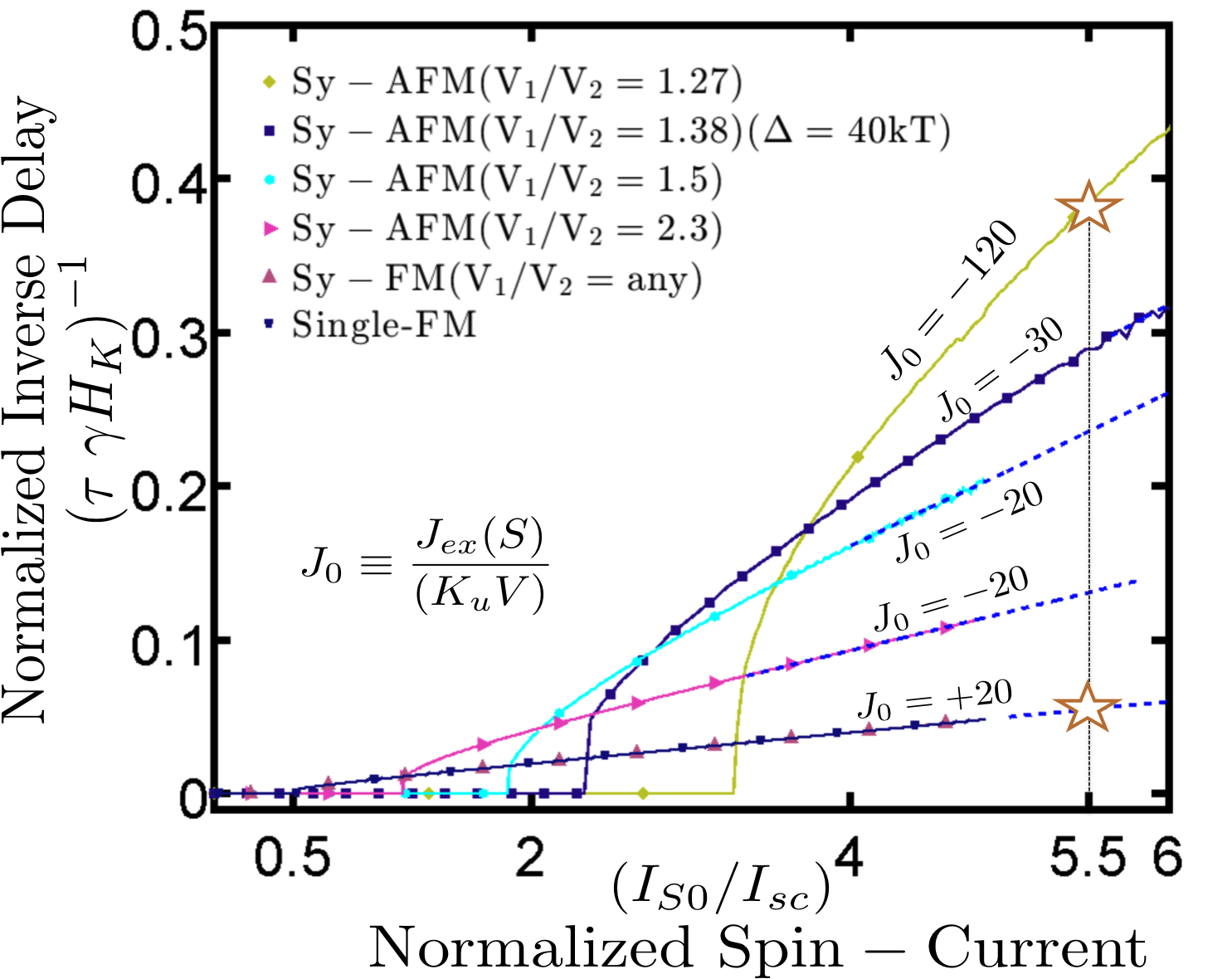}
 \caption{\textbf{Central Result}: The inverse delay (Eq.~(\ref{eq:delay})) normalized to $\gamma H_K$  is plotted as a function of the  spin-current  $I_{S0}$ normalized to the switching threshold of the single FM (FIG.~(\ref{fi:fig1})).  $I_{S0}$ is the current that is applied to one layer only, the total spin-current is 2$\ I_{S0}$. The slopes (dashed lines) for  high overdrive are exactly given by: $\mathrm{s}=(\tau H_K \gamma)^{-1}/(I_{S0}/I_{sc})$=$\alpha \left(N_1+N_2/N_{net}\right)$ where $N_i=(M_s \rm Vol.)/\mu_B$ and $N_{net}=(N_1-N_2)$ for the Sy-AFM and $N_{net}=N_1+N_2$ for Sy-FM and $N_{net}=N_1+N_2$ for the single FM. A normalized exchange interaction is defined, $J_0=J_{ex}(S)/K_u V$ where $J_{ex}$ is in units of ergs/$\rm cm^2$ and $K_u V=60$ kT.  The  parameters are $H_K^{\rm eff}=5000 $ Oe, the PMA diameter $\Phi=36$ nm, $M_s=1000 \rm emu/cc$ and a damping coefficient $\alpha=0.01$.   $I_{sc}$ is the  switching threshold of the single FM, and $\approx 15.4 \rm \ \mu A $. For the chosen parameters, $J_0=\pm 20$ corresponds to $\rm J_{ex}=\pm 5 \rm \ erg/cm^2$. For  a  $\Delta=40 \ \rm kT$ magnet, with identical magnetic properties ($M_s=1000 \rightarrow \ 650 \ \rm emu/cc$) normalized exchange becomes $J_0=-30$ for the same $\rm J_{ex}$. A value of $J_0 =-120$ corresponds to  $\rm J_{ex}=30 \ \rm erg/cm^2$, which is below the maximum strength experimentally measured in Co/Rh/Co structures.  Shorter delays are attainable when approaching the limit $\beta=1$, however a stronger exchange coupling would be required. For weaker exchange coupling, the magnets  might go to a meta-stable state until the pulse is turned off, see FIG.~(\ref{fi:fig5}). The thresholds for switching in all cases are given analytically in the text. The stars mark the switching conditions shown in FIG.~(\ref{fi:fig3}).}
\label{fi:fig2}
\end{center}
\end{figure}

It is commonly assumed that to achieve a given switching delay, the current has to exceed the critical current by a certain factor and therefore a higher critical current implies a higher switching current. However, we show that this is not true for SFM structures which can provide significantly reduced switching delay for a given current density, even though the critical current is increased in comparison to the ferromagnetic structures. The central result of this work is shown in FIG.~(\ref{fi:fig2}) which shows the inverse switching delay as a function of the spin current $I_{S0}$ applied to one of the layers, normalized to the critical switching current of the single FM layer in FIG.~(\ref{fi:fig1}c). The single FM and the synthetic FM have identical switching delays and switching thresholds since the constituent layers have identical material parameters. With sufficient exchange coupling strength, a synthetic FM behaves essentially as a single FM. The striking result, shown in the figure, is that the inverse switching delay increases at a much faster rate for the anti-ferromagnetic (AFM) configuration as the spin current is increased. In addition, the rate of increase strongly depends on the relative thickness of the constituent layers.

    This result is directly obtained from numerical simulations based on coupled Landau-Lifshitz-Gilbert (LLG) equations, and it can be understood from a simple angular momentum conservation argument that requires that the minimum current-delay product to be limited by the net number, $N$, of Bohr magnetons \cite{sun2000spin,behin2011switching} comprising the structure. A bilayer with $N_{1,2}=(M_s V)_{1,2}$ Bohr magnetons in layers 1 and 2, has a total of  $(N_1 + N_2)$ for FM coupling, and  $(N_1 -N_2)$ for AFM coupling. Consequently the slope of the inverse switching delay versus normalized current in FIG.~(\ref{fi:fig2}) equals $(N_1 + N_2)^{-1}$ for FM coupled layers and is larger, $(N_1 -N_2)^{-1}$, for AFM coupled nanomagnets.

    Note that the layers of the bilayer structure are chosen to have the same anisotropy field, $H_K$, and magnetization per unit volume, $M_s$. The difference in $N_1$ and $N_2$ arises simply from the difference in thickness or volume. This simple requirement is of significant practical importance, as it allows the synthetic ferrimagnet to be fabricated using the  \textit{same magnetic material}. There is no restriction on the magnitude of the material's magnetic anisotropy, except that low $H_K$ materials need to have  higher values of $N_{1,2}$ to ensure thermal stability, thereby resulting in worse current-delay products. For AFM-coupled bilayers, $(N_1 -N_2)$ can be made very small, even if $N_{1,2}$ are individually large. In principle  $(N_1 -N_2)$ can be made arbitrarily small, however this requires very large exchange coupling energies $\rm J_{ex}$. This requirement places a practical limit on how small $(N_1 -N_2)$ can be made. We note that the exchange coupling strengths given in FIG.~(\ref{fi:fig2}) are   experimentally demonstrated values in anti-ferromagnetically coupled structures via Ru/Rh interlayers \cite{parkin1991systematic,zoll1997giant,zoll1997preserved}, where the reported coupling strengths  range from  $\rm J_{ex}=1-34 \rm \ erg/cm^2$.
    
We would like to stress that our proposal of building synthetic ferrimagnets out of identical magnetic materials is very different from the well-established principle of coupling low to high magnetic anisotropy materials which are known to reduce the critical current, but at the expense of switching time delay because it increases $N_{1, 2}$ relative to the high anisotropy layer. Our proposal on the other hand leads to a slight increase in the critical current, but for a given current provides a striking reduction in delay.

Finally, we would like to note that the recent experimental demonstration of spin-current driven domain wall  motion involving PMA-based synthetic antiferromagnets \cite{yang2015domain} bears striking similarities to the principle we discuss here, especially considering the observation that  the domain wall velocity in these systems  shows a monotonic increase as the net magnetization in the SAF structure is decreased. This is exactly the same trend we observe in FIG.~(\ref{fi:fig2}).

\section{Main Results}
We assume that the magnetic layers of the synthetic ferrimagnet  are well-described by a macrospin model in the monodomain approximation, and that the mean-field approximation describes the exchange interaction between layers 1 and 2. The coupled LLG equation that is the basis of all results in this paper is given by:
\begin{eqnarray}
&& (1+\alpha^2)\frac{d\hat m_i}{dt} = -|\gamma|{\hat m_i \times \vec{H}_i} - \alpha |\gamma| (\hat m_i \times \hat m_i \times \vec{H}_i)\nonumber \\  && +  \frac{1}{q  N_i}(\hat m_i \times \vec{I}_{Si} \times \hat m_i)  + \left(\frac{\alpha}{q N_i} (\hat m_i \times \vec{I}_{Si})\right)
\label{eq:llg}
\end{eqnarray}
\noindent where $i$ stands for magnets $1,2$ respectively. Each magnet is assumed to have perpendicular magnetic anisotropy (PMA), therefore the effective field including the mean-field exchange component can be written as: $\vec{H}_i= H_{K} m_{zi} \  \hat z + J_{ex}(S_i+S_j)/(M_s V)_i \ \hat m_j$ where $i, j \in \{1,2\}$, $i \neq j$,  $N_i = (M_s V)_i / \mu_B$, and $\mu_B$ is the Bohr magneton. We define $H_K$ as the effective perpendicular anisotropy that is the net difference between the surface and shape anisotropy, i.e  $H_K= H_K^{eff}=H_K^s-4\pi M_s$, throughout this paper. $S_{i,j}$ is the surface area of the layers. The spin current inputs $\vec{I}_{Si}$ are applied along the $+z$ direction,  are assumed to be of equal magnitude and are present throughout the entire magnetization reversal time. This is referred in this paper as  static current switching in contrast to pulsed current switching. Numerical parameters used for the results of FIG.~(\ref{fi:fig2}) are: $H_K = 5000 \rm \  Oe$,  $M_s =1000 \rm \ emu/cc $, PMA diameter $\Phi=36 \rm \ nm$, damping coefficient $\alpha=0.01$ and $t_1+t_2= 1 \rm \  nm $. The thermal stability for the system is $\Delta = 60 \rm \ kT$ assuming that it is  given by the sum of the thermal stability of the constituent layers. A value of  $\rm J_{ex}$= $\pm 5 \ \rm erg/cm^2$, is chosen, as measured experimentally in Co/Ru/Co multilayers \cite{parkin1991spin}. The $x-$axis in FIG.~(\ref{fi:fig2}) is normalized to $I_{sc}= 4\ q/\hbar \  \alpha \  ( \Delta)$ the critical switching current required for the single FM layer, and is equal to $I_{s0}\approx 15.4 \rm \ \mu A $. For the chosen PMA diameter ($\Phi= 36 \rm \  nm$), this corresponds to a critical current density of $J_c\approx 2\times 10^5 \rm \ A/cm^2$. 

\subsection{Delay Definition}
\begin{figure}[b]
\begin{center}
\includegraphics[width=\linewidth]{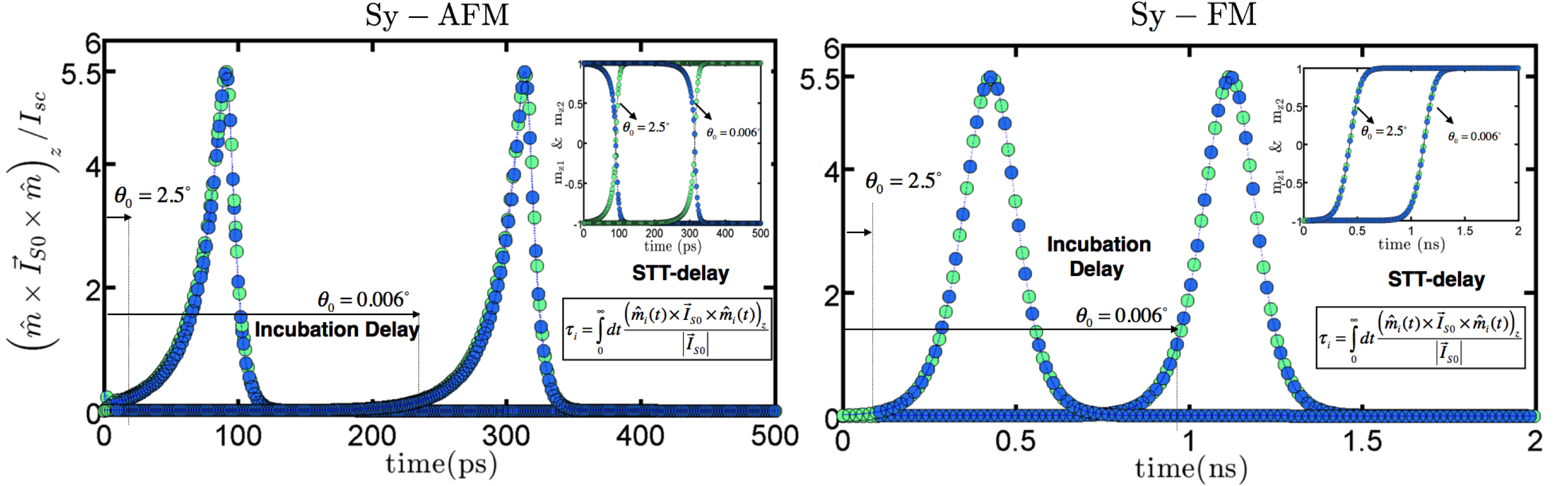}
\caption{\textbf{Delay defined as independent of  initial angle, $\theta_0$:} The z-component of spin-torque currents of both layers $\hat m \times \vec{I}_{s0} \times \hat m$ for synthetic AFM and synthetic FM for a fixed overdrive (Marked in FIG.~(\ref{fi:fig2})) are plotted for two different initial angles $\theta_0=2.5^{\circ}$ and $\theta=0.006^{\circ}$. Note that the incubation delay (given by the horizontal arrows in the figures) is longer for the smaller initial angle $\theta_0$, but \textbf{the spin-transfer-torque delay is  independent of $\theta_0$}, since it is  defined in terms of the area under the pulse (See Eq.~(\ref{eq:delay})) which remains approximately constant. The relative increase in the speed for the synthetic AFM can  be observed by the reduced width of the Gaussian-shaped spin-torque currents which signifies a reduced angular momentum. Insets show the magnetization transients for individual layers for synthetic FM and AFM. In the case of synthetic FM, $m_{z_{1}(t)}$ and $m_{z_{2}}(t)$  are identical, therefore appear as a single line.}
\label{fi:fig3}
\end{center}
\end{figure}

It is known that the  magnetization delay is a strong function of the initial angle of the magnets \cite{sun2000spin} and in tilted media,  magnetic layers are engineered to have built-in initial angles  to increase their switching speed \cite{zou2003tilted}.  In this paper, we define delay in terms of angular momentum transfer that is independent of the  chosen initial angle as shown in FIG.~(\ref{fi:fig3}). The switching delay for a given layer $\tau_i$ is the  ratio of deposited charge  ($Q$) to the spin-current input ($|\vec I_{S0}|$) applied to that layer: 

\begin{equation}
\tau_{i}    = \frac{Q}{\left| \vec I_{S0} \right|} = \int \limits_0^\infty  {dt\frac{{{{\left( {\hat m_i(t) \times {{\vec I }_{S0}} \times \hat m_i(t)} \right)}_z}}}{{\left| {{{\vec I }_{S0}}} \right|}}} 
\label{eq:delay}
\end{equation}

The integrand in Eq.~(\ref{eq:delay}), the z-component of the spin-torque current, is shown as a function of time during switching in FIG.~(\ref{fi:fig3}) for synthetic FM and AFM layers. The integral is simply the area under the spin-torque current and is approximately zero throughout the incubation delay; thereby making the delay  independent of the chosen initial angle.

In the near to high overdrive current regimes, the area under the spin-torque current (summed for layer 1 and layer 2) yields exactly the net number of magnetic moments in the bilayer, which is $2 (N_1+N_2)$ for synthetic (and single) ferromagnets and $2 (N_1-N_2)$ for synthetic ferrimagnets. It has been noted in \cite{behin2011switching}, the integrand of Eq.~(\ref{eq:delay}) is exactly equal to $2 N$ for single (PMA) magnets, however this is strictly true only in the high-overdrive regime (See Supplementary Information).  As the overdrive is increased, the time-integral of Eq.~(\ref{eq:delay}) behaves as a Gaussian: increasing in maximum amplitude, but becoming narrower in order to keep the area underneath constant, a  manifestation of angular momentum conservation. Therefore, when delay is defined as in Eq.~(\ref{eq:delay}), the inverse delay becomes exactly proportional to the net number of spins in the system as shown by the dashed lines in FIG.~(\ref{fi:fig2}). 

Consider next the figure of merit, $E\times\tau$, namely the product of the switching energy and the switching delay: As shown in \cite{sarkar2014charge}, this metric can  equivalently  be expressed by the static parameters of the total deposited charge over a given resistance, i.e  $E\times\tau = Q^2 R \ (\tau_{sw}/\tau_{pw})$ , where \textit{Q} is the charge deposited into the system and \textit{R} is the net resistance that the injected current experiences, $\tau_{sw}$ and $\tau_{pw}$ are the switching delay of the magnetization reversal and the pulse duration of the applied spin-current respectively. In this paper, most of the results presented are for $\tau_{sw}$ = $\tau_{pw}$ since the spin-currents are assumed to be on during the entire magnetization reversal time. Therefore, the net charge ($Q$) required for switching is reduced, improving the energy-delay requirements for the nanomagnets significantly \cite{nikonov2015benchmarking}.

\subsection{Symmetric Currents}

\begin{figure}[b]
\begin{center}
\includegraphics[width=\linewidth]{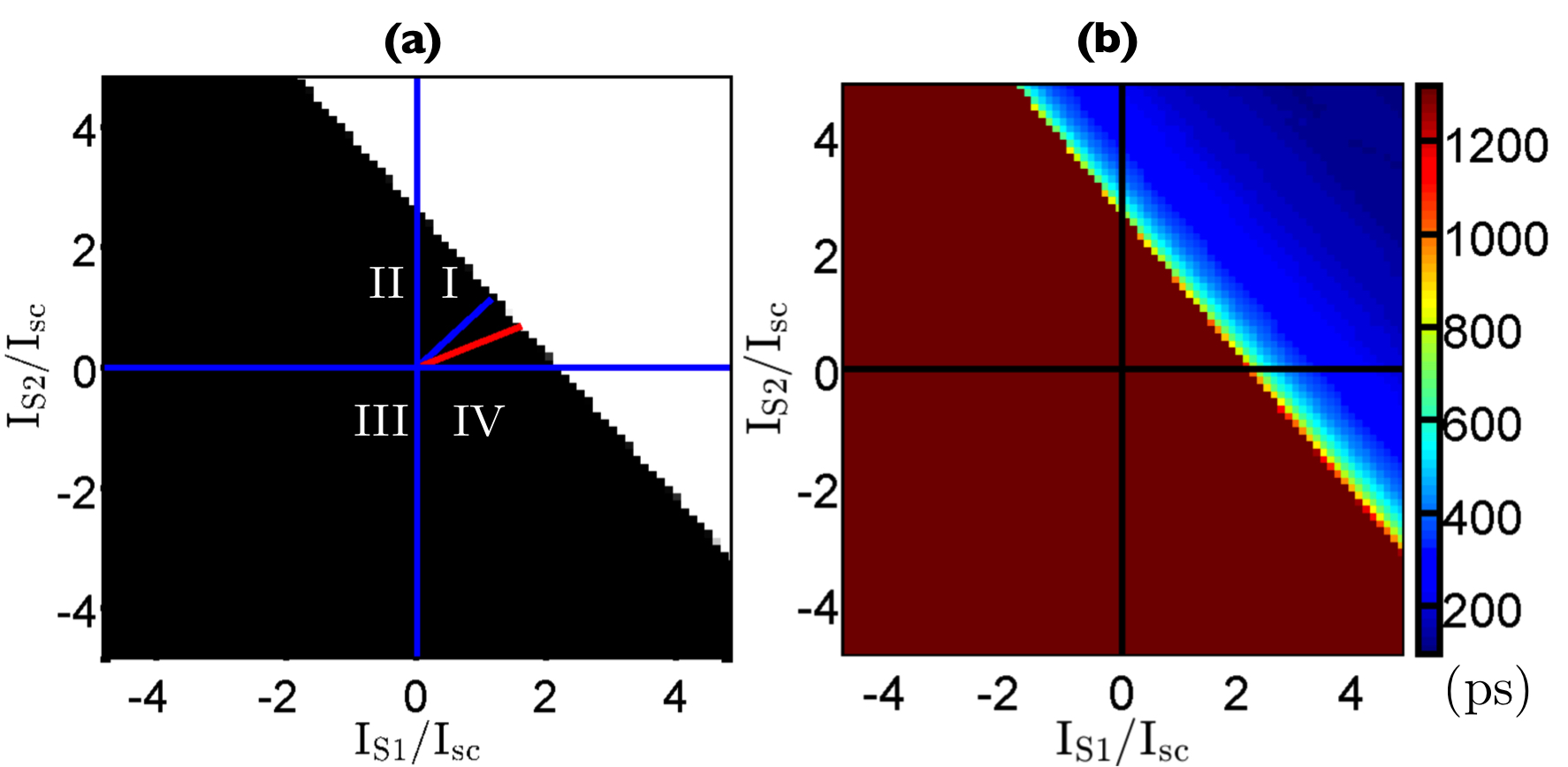}
\caption{\textbf{Phase plot}: (a) We show all four quadrants for layer 1 and  layer 2 currents, emphasizing that the optimum quadrant is I, where spin-currents to both layers are in the same direction. Currents are normalized to $I_{sc}$ as in FIG.~(\ref{fi:fig2}). The initial condition in all cases is assumed to be $(-1,1)$. The black regions correspond to no switching ($mz_1=-1) \ (mz_2=+1)$  and the white region corresponds to switching ($mz_1$=$+$1,\ $mz_2$=$-$1). The red line indicates the minimum amount of dimensionless spin current: Equal for both layer 1 and layer 2, such that $\vec{I}_{S1}/ (\gamma q N_1)=\vec{I}_{S2}/ ( \gamma q N_F)=I_{0} \hat z$, for this example $V_2/V_1=3/7$. The blue line is obtained by assuming that the total sum of applied spin current is invariant for rigid coupling ($\rm J_{ex}=-5 \ erg/cm^2). $ (b) Phase plot for switching delay (based on Eq.~(\ref{eq:delay})). The color bar is scaled to picoseconds. Dark red regions correspond to no switching. }
\label{fi:fig4}
\end{center}
\end{figure}

One of the non-intuitive aspects of the synthetic ferrimagnet free layer we propose, is the requirement of symmetrically spin-polarized currents to be applied to both layers, counter-intuitive for the AFM configuration. If the exchange interaction was weak,  we would naturally expect to apply anti-symmetrically polarized spin-currents to the layers to switch their orientation. However, we observe that in the case of rigid coupling, anti-symmetrically polarized currents are much less efficient than symmetrically polarized currents as shown in FIG.~(\ref{fi:fig4}). This is also reflected in the analytical switching thresholds we derive in the next section. 

We observe that the optimum current configuration is when the spin-current polarization applied to the thicker layer ($t_1>t_2$) is in the ``correct'' direction for switching, i.e in the anti-parallel direction to its original direction while the spin-current applied to layer 2 is in the ``wrong'' direction, i.e. aligned parallel to its own magnetization, which would normally not cause switching if the layers were decoupled ($J_{ex}=0$). FIG.~(\ref{fi:fig4}) provides a phase plot showing  four quadrants for combinations of spin-currents {$I_{S1}$ and $I_{S2} $}  that are applied along the z-axis of the AFM coupled magnetic layers, showing the first quadrant to be the optimum region. Critical switching threshold current values (FIG.~\ref{fi:fig4}a) and magnetization delays (FIG.~(\ref{fi:fig4}b)), are shown, the latter being calculated based on Eq.~(\ref{eq:delay}). To exclude dynamic effects associated with the input pulse duration and shape,  we employ static currents in obtaining the results of FIG.~(\ref{fi:fig4}), similar to our scheme in FIG.~(\ref{fi:fig2}). The numerical parameters are chosen to be the same as FIG.~(\ref{fi:fig2}), with $V_1/V_2=3/7$.

\subsection{Analytical Results}

We derive analytical formulas for critical switching threshold (the x$-$intercepts in FIG.~(\ref{fi:fig2})) by linearizing the LLG equation around the fixed points of the dynamic AFM and FM systems for various limits (See Supplementary Information for detailed derivations). The single FM case for a PMA magnet is well-known, thus we address the  FM bilayers first. Analytical work for exchange coupled in-plane magnets (IMA) has been conducted by others for bilayers that are driven by a single spin-current source \cite{you2010critical,balavz2013current,lacoste2014magnetization,koop2014quantitative}. Our approach differs in two ways: (a) We focus on synthetic bilayers driven by two distinct spin-currents, symmetric or anti-symmetric in spin-polarization direction; and (b) We focus on identical PMA materials with the only asymmetry being their difference in thicknesses. Therefore, we obtain  simple expressions that to the best of our knowledge have not been previously reported.

For strongly exchange coupled FM-bilayers having equal $K_u$ and $\alpha$ and $M_S$ parameters and differing only by volume, the sum of the critical switching currents can be shown to be equal to the sum of spin-torque switching currents of the individual (decoupled) magnets.
\begin{equation}
I^c_{Si} = \frac{2 q}{\hbar}\alpha   \left(K_i V_i \right)
\label{eq:singlefm}
\end{equation}
where $i \in \{1,2\}$ and $I_{S1}$, $I_{S2}$ are the minimum spin currents applied to layers 1 and 2 respectively.  We derive this result by a Jacobian analysis assuming equal dimensionless spin-currents being applied to the individual layers i.e, $I_{S1}/q N_1 =  I_{S2}/ q N_2$ and find that this result is independent of the exchange strength $\rm J_{ex}$ (Supplementary Information). However, numerical simulations suggest that as long as the total spin-current  given by  $I^c_{S1}+I^c_{S2}$ in Eq.~(\ref{eq:singlefm}) is split in half and applied equally to each layer ($I_{S_{1,2}}=I_{S0}=(I^c_{S1}+I^c_{S2})/2$), the magnets switch without requiring equal dimensionless current. Eq.~(\ref{eq:singlefm}) is intuitive since one would expect a rigidly coupled synthetic-FM to behave like a single FM with an effective total $K_u V$. 

Next, we consider a synthetic ferrimagnet comprising two FM layers antiferromagnetically coupled and having unequal volumes ($V_1 > V_2$) which are driven by symmetrically polarized spin-currents, all magnetic layer parameters are assumed to be equal otherwise. The derivation for this case also assumes equal dimensionless spin-currents applied to both layers,  however as the phase plot in FIG.~(\ref{fi:fig4}) shows for strongly exchange coupled AFM structures, the total current required to switch the synthetic ferrimagnet does not depend on the individual proportion of the injected spin-currents,	$I_{S1}$ and $I_{S2}$. It can be observed from the phase plot that the bilayer can be switched as long as the sum of  $I_{S1}$  and $I_{S2}$ equal a constant value. The critical spin current  that needs to be applied to layer 2 is:

\begin{equation}
{I^c_{S2}} = \frac{{{(\beta  - 1}){I_{ex}}}}{2} \hspace{-0pt}+\hspace{-0pt} \sqrt {\hspace{-0pt}I_c^{}\left( {{I_c} + {I_{ex}}\left( {\beta  \hspace{-0pt}+ \hspace{-0pt}1} \right)} \right)\hspace{-0pt} +\hspace{-0pt} \frac{{I_{ex}^{2}}}{4}{{\left( {\beta \hspace{-0pt} -\hspace{-0pt} 1} \right)}^2}} 
\label{eq:AFMc}
\end{equation}

where $\beta=V_2/V_1 \le 1$ and $I_c$ and $I_{ex}$ are defined as:
\begin{equation}
{I_{ex}} = \displaystyle\frac{{2q}}{\hbar }\alpha \left( \rm |{{J_{ex}}|S} \right)
\label{eq:exx}
\end{equation}
\begin{equation}
{I_c} =  \displaystyle\frac{{2q}}{\hbar }\alpha \left( {2{\rm{ }}{K_2}{V_2}} \right)
\label{eq:sun}
\end{equation}

The current that needs to be applied to layer $V_1$, assuming an equal dimensionless spin-current ($I_{S1}/q N_1 =  I_{S2}/ q N_2$), is $I^c_{S1}=I^c_{S2}/\beta$. This result is exact and works for all values of $\rm J_{ex}$ from weak to strong exchange coupling. We have, however, confirmed by numerical simulations that for strong coupling, the total spin-current is given approximately by the sum of $I^c_{S1}+I^c_{S2}$, as shown in FIG.~(\ref{fi:fig4}). Then, for the case of a symmetric structure in which equal total currents are provided to both layers,  the minimum input currents  become: $I_{S1}=I_{S2}\approx I^c_{S2} (1+1/\beta)/2$. The red line in FIG.~(\ref{fi:fig4}) shows the analytical threshold which yields a higher layer 1 current since $N_1  > N_2$ for equal dimensionless currents, and the blue line is obtained by halving the total spin-current necessary. 

Note that the equal volume case ($\beta=1$) imposes an upper limit to the threshold current:
\begin{equation}
\lim_{\beta\rightarrow 1}I^c_{S0}  =  \sqrt{(I_c^2 + 2 I_c I_{ex})} \quad \quad \mbox{\rm (Symmetrically driven)}
\label{eq:afm}
\end{equation}
where $I^c_0$ is the spin-current applied to both layers, since layer 1 and 2 are identical in this limit. Eq.~(\ref{eq:afm})  shows that even when the exchange interaction is  large compared to the uniaxial anisotropy constant ($I_{ex} \gg I_c$), the critical current ($I^c_{S0} \approx \sqrt{2 I_c I_{ex}} $) does not diverge, on account of the square root dependence. This is in sharp contrast with the case where anti-symmetrically polarized spin currents are applied to the  layers of the AFM, i.e  $I_{S2}=-I_{S1}$.  In this case, we show  that the critical  current that needs to be applied to both layers to create an instability off the equilibrium points is (assuming $\beta=1$):
\begin{equation}
\label{eq:afm_asym}
I_{S0}^c = \left( {{I_{ex}} + {I_c}}\right) \quad \quad \mbox{\rm (Anti-symmetrically driven)}
\end{equation}
where $I_{ex}$ and $I_c$ are given by Eq.~(\ref{eq:exx}) and Eq.~(\ref{eq:sun}) respectively. Eq.~(\ref{eq:afm_asym}) shows that in the case of anti-symmetric currents, the deviation threshold grows linearly as a function of the exchange interaction, $\rm J_{ex}$, and therefore it becomes large when compared to Eq.~(\ref{eq:afm}). This behavior is also confirmed by the phase plot shown in FIG.~(\ref{fi:fig4}), in which the IV quadrant corresponds to the bilayer being driven by anti-symmetrically polarized spin currents.

\subsection{Effects of exchange strength, $\rm |J_{ex}|$}
\begin{figure}[b]
\begin{center}
\includegraphics[width=\linewidth]{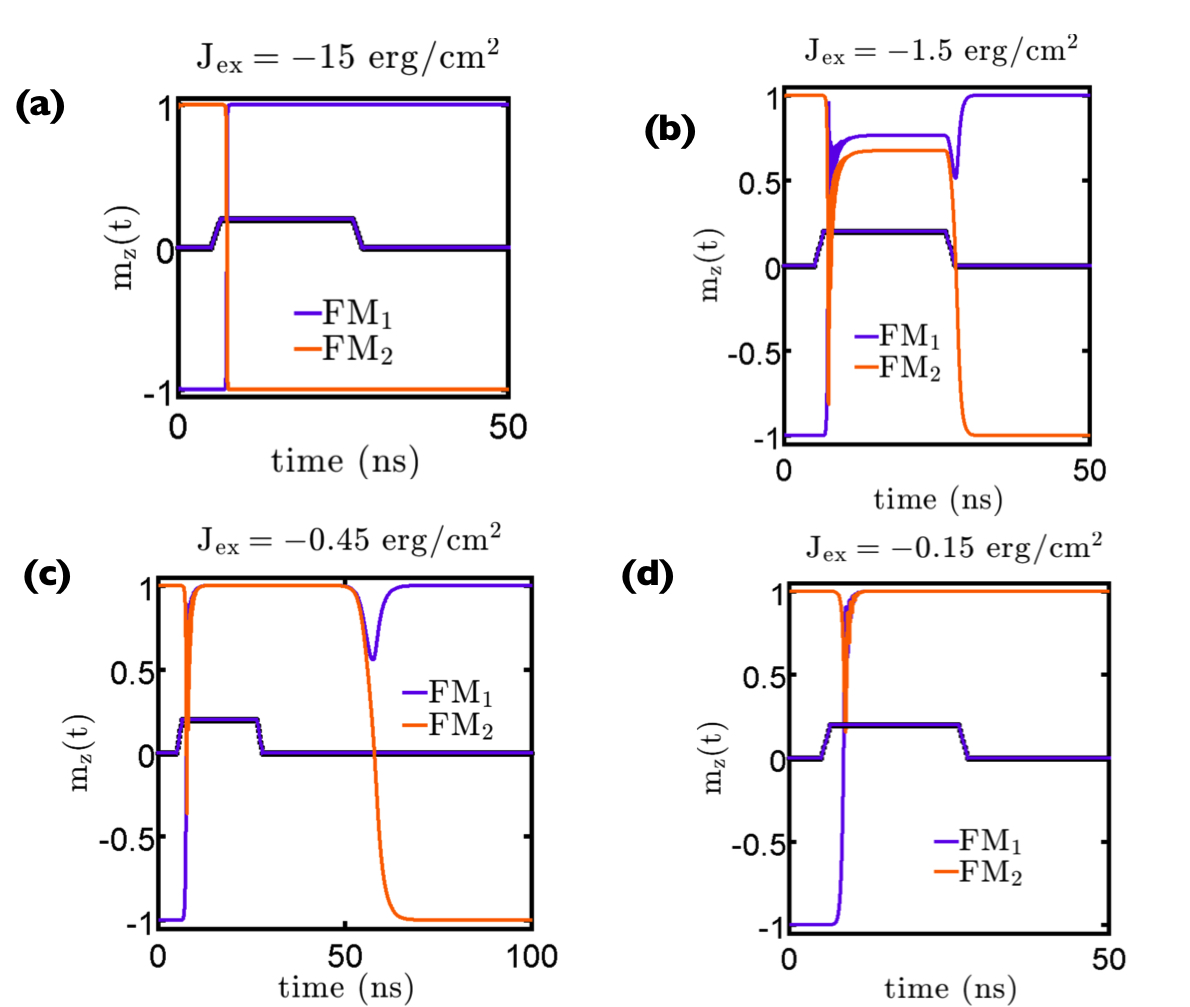}
    \caption{\textbf{Effects of exchange strength, $|J_{ex}|$}: Examples of switching simulations of the FM layers comprising the synthetic ferrimagnet are shown, for different values of the exchange coupling strength and a constant write current pulse width. A diameter of $\Phi=100$ nm and $\Delta=60$ kT is chosen for illustration. The switching mechanism works in all cases except for the very weak coupling case shown in (d) where the exchange energy is so weak that even after the pulse is turned off,  the bilayers do not form an AFM configuration.}
\label{fi:fig5}
\end{center}
\end{figure}

The fact that we apply a constant spin-current to the thinner layer in the direction to pin this layer in its initial  state might cause switching errors if the exchange interaction is not strong enough.   In FIG.~(\ref{fi:fig5}) we investigate this behavior, and show the response of the layers' magnetization as a function of time upon applying a rectangular current pulse. FIG.~(\ref{fi:fig5}a) shows the case for strong coupling ($\rm J_{ex}=-15 \ erg/cm^2$) in which the switching behavior does not depend on the duration of the pulse, as the exchange interaction is strong enough to keep the individual layers anti-parallel at all times. Upon the spin-currents attaining their peak value, the magnets switch and remain in their switched configuration. FIG.~(\ref{fi:fig5}b) shows an example of moderate coupling ($\rm J_{ex}=-1.5  \ erg/cm^2$).  In this case, since the applied current is much larger than the individual critical current of the thinner layer, the system  reaches a meta-stable state as long as the pulse is on, but when the pulse is turned off, the layers ultimately go to a $(-1,+1)$ state starting from a $(+1,-1)$ state, completing the correct reversal. FIG.~(\ref{fi:fig5}c) shows the case for weak coupling ($\rm J_{ex}=-0.45  \ erg/cm^2$),  the thicker layer switches once the peak value of the current pulse is attained. However, the thinner layer 2, remains pinned in its original direction due to the parallel orientation of the spin-current with the initial direction of its magnetization. After the pulse is turned off, the anti-ferromagnetic exchange interaction is strong enough to keep the layers in an anti-parallel state. This example shows that even for weak coupling, short pulses can be used to switch the magnets correctly as long as the exchange interaction is strong enough to force an AFM configuration in equilibrium. The final example, FIG.~(\ref{fi:fig5}d) shows how in the case of very weak coupling ($\rm J_{ex}=-0.15  \ erg/cm^2$) a switching failure ensues. The switching of the thicker layer completes at the peak of the current pulse, and remains switched long after the current pulse is turned off. However, the exchange interaction is not strong enough to force an AFM configuration.
 
Therefore at sufficiently high exchange coupling energies between the layers and using short current pulses, one might obtain even faster switching times as $\beta\rightarrow1$ as shown in FIG.~(\ref{fi:fig2}). 

In summary, FIG.~(\ref{fi:fig5}) shows that the symmetrically polarized spin-current reversal mechanism here discussed works even when the exchange interaction is in the moderate-to-weak regime and successful reversals can be  achieved utilizing short current pulses.

\subsection{Double Fixed Layer MTJs}

\begin{figure}[b]
\begin{center}
\includegraphics[width=4in]{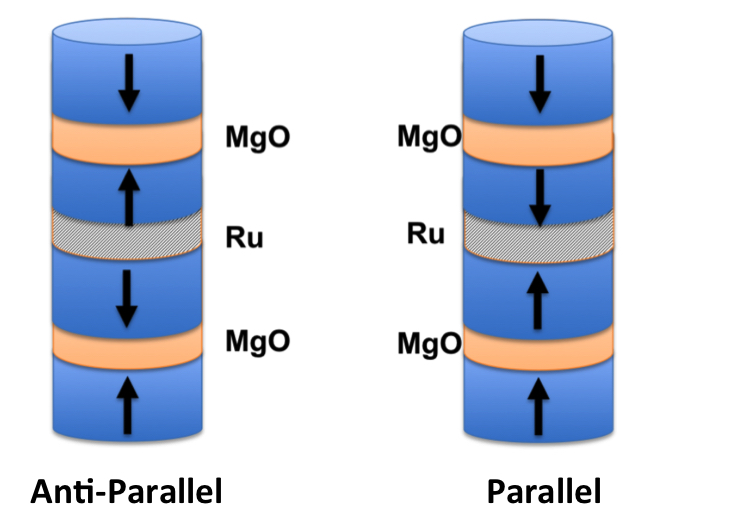}
\caption{\textbf{Proposed device implementation}: An MTJ-based magnetic architecture of the symmetrically current driven synthetic ferrimagnetic structure. Double fixed layers with anti-parallel magnetizations provide independent spin currents to layers 1 and 2 for corresponding injected charge currents. Assuming that the Ru interlayer  separates the spin-conductance between the the top and the bottom, the  full structure becomes a series combination of two Parallel or Anti-Parallel MTJs. }          
\label{fi:fig6}
\end{center}
\end{figure}

One specific implementation of the synthetic ferrimagnet described in this work is given in FIG.~(\ref{fi:fig6}). The efficiency of the switching mechanism we have described  increases when  two distinct spin-currents are applied. In principle, they can be supplied from other spin-current sources, such as the Giant Spin Hall Effect (GSHE) in a 3-terminal device configuration \cite{liu2012spin} which we do not discuss further in this paper. To generate independent spin-current inputs in the same direction, the architecture shown in FIG.~(\ref{fi:fig6}) which employs two reference layers aligned in anti-parallel directions is proposed.  

The use of double-reference layers in MTJ single-free layer structures to increase the spin-torque efficiency by repolarizing the charge current while it exits the MTJ structure, was first pointed out by Berger \cite{berger2003multilayer}, and several experiments have been performed to validate the concept \cite{fuchs2005adjustable,meng2006low,cuchet2015perpendicular,clement2015modulation}.

Synthetic antiferromagnets often employ Ru \cite{parkin1991spin} as an exchange coupling interlayer whose thickness is adjusted to achieve the desired level of exchange strength between the layers. In our current treatment we assumed that the magnetic layers are effectively driven by independent spin-currents disregarding the transport effects throughout the structure. This would be an accurate assumption if Ru acted as an ideal spin-sink, however, Ru has a spin-flip length of $\lambda_{sf} \approx 14 \ \rm nm$ \cite{eid2002current} that is much longer than the typical spacer thicknesses ($t_{Ru}\approx 0.3 - 0.6 \rm \ nm$), and may not be an ideal spin-sink. In that case a detailed transport model is needed, similar to the treatment  in \cite{hernandez2010calculation}. Such a treatment is beyond the scope of this paper.

Structures similar to the one shown in FIG.~(\ref{fi:fig6}) have been shown to exhibit TMR values comparable to standard MTJs   \cite{raychowdhury2011numerical,ghosh2014micromagnetic}. Note that in our proposal, this is due to the counter-intuitive necessity of requiring symmetric spin-currents to be applied to both magnetic layers.

\subsection{Device Considerations}

\subsubsection{Switching Current Limits}
One of the critical design parameters for STT-MTJ devices is the need to restrict the injected charge current density to magnitudes below the dielectric breakdown of the MgO tunnel barrier layer. Whereas different values for the voltage breakdown for nanoscale MgO layers have been reported, there is general consensus that the breakdown voltage is around $\ \approx 0.4  \ \rm  V $ \cite{min2010study}. The current densities required to switch the synthetic ferrimagnet here proposed are shown to be below this MgO breakdown constraint. Consider MTJ cylindrical stacks of $\Phi= 36 \rm \  nm $ in diameter comprising PMA magnets having effective  magnetic anisotropies of $H_K=5000 \ \rm Oe$ and the magnetic properties employed for the results of  FIG.~(\ref{fi:fig2}). Assuming an RA-product of $\rm 4 \  \rm \Omega-\mu m^2$ \cite{yakushi2010} and a $400 \rm \  mV$ breakdown voltage \cite{min2010study}, the breakdown current becomes $ 100 \ \rm \mu A$.  Assuming a  polarization factor of $P \approx 0.5$ \cite{yakushi2010}, we conclude that the overdrive currents employed in FIG.~(\ref{fi:fig2}) are below the current breakdown limit for MgO. 

\subsubsection{Ultra-fast Switching: Low/High $K_u$ }
As shown in FIG.~(\ref{fi:fig2}), the switching speed of synthetic ferrimagnets is ultimately determined by the net number of spins in the system. This means that the employment of very high $H_K$ magnets is not a necessary condition for exploiting their benefits,  provided that the bilayers are strongly exchange coupled. For example if low $H^{eff}_K = \rm 100 \ Oe$ PMA magnets are chosen, to meet a $\Delta = 60 \rm \ kT $ thermal stability criterion, the pillar diameter and the layer thicknesses need to be increased to ($\Phi = 100 \rm nm$) and ($t_1 = 4 \rm nm$ and $t_2 = 3 \rm \ nm$) respectively.  Numerical simulations estimate that for an approximately $0.3 \rm \  mA$ spin-current applied to both layers (well below the breakdown requirements in this larger area), sub-nanosecond switching delays are attainable even with such a low $H_K^{eff}$.

Next we consider a state-of-the-art MTJ nano-pillar dimension of $\Phi=20 \rm \ nm$ for a storage density of $\approx 1 \rm Tb/{in}^2$\cite{sato2014properties}. The thickness of the AFM coupled layers is selected to be $t_1= 4 \rm \  nm $ and $t_2= 3 \rm \  nm $. The use of thicker PMA magnets allows precise thickness control,  facilitating fabrication and reliability of devices approaching the equal thickness regime. The following   magnetic parameters for both layers are chosen in our estimate to provide a bilayer thermal stability of $\Delta = 55 \rm \ kT$, $M_s =210 \rm \ emu/cc $ with  $H_K = 1 \rm \  T$ with $\alpha=0.01$,  for both layers. We note that these magnetic properties are readily met in materials currently employed in the fabrication of hard drives. For this example, a rigidly coupled AFM system requires $N_{net} =N_1-N_2 \approx 3500$ $\mu_B$. Using a static spin current of 100 $\mu A$,  numerical simulations show that a delay that is of the order of picoseconds $\approx 12.8$ ps  (calculated from Eq.~(\ref{eq:delay})) is attainable, provided that the exchange interaction between the layers is around ($\rm J_{ex}=-40  \ erg/cm^2$).

\section{Conclusions}
We have shown that strongly coupled synthetic anti-ferromagnets can potentially switch significantly faster than synthetic or natural ferromagnets having the same thermal stability. This is achieved by circumventing the conventional angular momentum constraints to switch single of synthetic ferromagnets, whereby the number of injected spin polarized electrons needs to equal the total number of Bohr magnetons comprising the system. The synthetic ferrimagnet here proposed, decreases the total  angular momentum needed to be deposited on the magnets via the spin-transfer-torque mechanism. In principle, the angular momentum can be reduced to zero in compensated ferrimagnets employing equal volume layers, however, this requires large exchange coupling strengths, thereby imposing a practical limit to the realization of said compensated ferrimagnets. One of the most salient attributes of the solution here described is its simplicity: the magnetic layers comprising the synthetic ferrimagnet, differ only in their volumes, this enormously simplifies their fabrication and can be expected to result in significant device improvements in reproducibility and reliability. We  provide analytical results supported by numerical solutions of the LLG equation describing the critical switching threshold of these structures, as well as possible implementations in modern MTJ stacks. These results can potentially lead to the start of a \textit{synthetic} anti-ferromagnetics based spintronics.

\section*{Acknowledgement}
This work was supported in part by C-SPIN, one of six centers of STARnet, a Semiconductor Research Corporation program, sponsored by MARCO and DARPA and in part by the National Science Foundation through the NCN-NEEDS program, contract 1227020-EEC. RF was supported by the Nanoelectronics Research Initiative through the Institute for Nanoelectronics Discovery and Exploration (INDEX) Center.

\section*{Author contributions statement}

KYC, AZP, RF, EEM, SD participated in conceiving the idea,  analyzing and reviewing the results.  KYC, EEM, SD wrote the manuscript. 
 
\section*{Additional information}

\textbf{Competing financial interests:} Authors declare no competing financial interests.

%
\clearpage
\appendix
\section{Derivation of Analytical Results}
Here we show the  derivations of Eq.~(\ref{eq:singlefm},\ref{eq:AFMc},\ref{eq:afm},\ref{eq:afm_asym}) of the main paper. Our starting point is Eq.~(\ref{eq:llg}) which is reproduced below:
\begin{eqnarray}
&& (1+\alpha^2)\frac{d\hat m_i}{dt} = -|\gamma|{\hat m_i \times \vec{H}_i} - \alpha |\gamma| (\hat m_i \times \hat m_i \times \vec{H}_i)\nonumber \\  && +  \frac{1}{q  N_i}(\hat m_i \times \vec{I}_{Si} \times \hat m_i)  + \left(\frac{\alpha}{q N_i} (\hat m_i \times \vec{I}_{Si})\right)
\label{eq:sup}
\end{eqnarray}
\noindent where $i$ stands for $1,2$ representing  layer 1 and layer 2  ($t_1 > t_2$)  respectively. Each magnet is assumed to be perpendicularly polarized (PMA) therefore the effective field that includes the mean-field exchange component can be written as: $\vec{H}_i= H_{K} m_{zi} \  \hat z + J_{ex}(S_i+S_j)/(M_s V)_i \ \hat m_j$ where $i, j \in \{1,2\}, i \neq j$,  $S_{i,j}$ is the surface area of the layers and we define as the ``effective'' anisotropy such that $H_K = H_K^{eff}=H_K^s - 4\pi M_s$,  denoting the net difference between the surface and shape anisotropy. $N_i=(M_s V)_i / \mu_B$ where $\mu_B$ is the Bohr magneton. 

In all of our results below the spin-currents that are applied to both layers are assumed to be equal per spin,  i.e $\vec{I}_{S1}/ q N_1=\vec{I}_{S2}/q N_2$ with  spin-polarization in the $\pm z$ direction. Our general approach is to perform a Jacobian analysis such that
\[ \frac{d\hat \delta}{dt }=\underbrace{\left. \frac{\partial f}{\partial \hat m } \right |_{_{\hat m=\hat m_0}}}_{\mathbf{J}} \hspace{-20pt}\hat \delta \]
where Eq.~(\ref{eq:sup}) is viewed in the form $({\hat m})^{'}=\mathbf{f}(\hat m)$ and $\delta=\hat m - \hat m_0$, $\hat m_0$ being a stable point of the system. In all three cases we calculate the Jacobian matrix for a specified initial condition and spin-current input  to investigate the real part of the eigenvalues for a stability analysis.  The Jacobian matrix in all  cases reduces to a $4\times 4$ matrix, with zeros in the full $6 \times 6$ matrix, due to the 2D nature of each LLG equation, with the reduced Jacobian:
\[J_r= \left[\begin{array}{cc} \displaystyle\frac{\partial f_x}{\partial m_x} & \displaystyle\frac{\partial f_x}{\partial m_y} \\ \\ \displaystyle\frac{\partial f_y}{\partial m_x} & \displaystyle\frac{\partial f_y}{\partial m_y} \end{array}\right]
\]

All our results  agree with the exact solution of Eq.~(\ref{eq:sup}) however we note that in some cases the criticality thresholds may not always give rise to switching, as expected from a linear stability analysis. 

\textbf{Synthetic FM (Eq.~(\ref{eq:singlefm}))} \\
Let $V_2/V_1 = \beta < 1$ for the constituent layers and  all other magnetic parameters be equal.  Assuming $\vec{I}_{S1}/ (\gamma q N_1)=\vec{I}_{S2}/ ( \gamma q N_2)=I_0 \hat z$ with the initial condition $m_{z1,2}=-1$,  it is convenient to define $j_{ex}= J_{ex}(S)(t_i+t_j)/(M_s V)_2$, so that the exchange fields on the individual layers are given by  $\vec{H}^{ex}_{2}=  j_{ex} \hat{m}_1$ and $\vec{H}^{ex}_{1} = \beta j_{ex} \hat m_2$, where $j_{ex}$ is a positive number for FM coupling.  The reduced Jacobian under these conditions reads:

\begin{widetext}
\[J_r= \left[ \begin {array}{cccc} -\beta\,j_{{{\it ex}}}\alpha-{\it I_0}-
\alpha\,H_{{K}}&-\beta\,j_{{{\it ex}}}-H_{{K}}+\alpha\,{\it I_0}&\beta
\,j_{{{\it ex}}}\alpha&\beta\,j_{{{\it ex}}}\\ \noalign{\medskip}\beta
\,j_{{{\it ex}}}+H_{{K}}-\alpha\,{\it I_0}&-\beta\,j_{{{\it ex}}}\alpha
-{\it I_0}-\alpha\,H_{{K}}&-\beta\,j_{{{\it ex}}}&\beta\,j_{{{\it ex}}}
\alpha\\ \noalign{\medskip}\alpha\,j_{{{\it ex}}}&j_{{{\it ex}}}&-
\alpha\,j_{{{\it ex}}}-{\it I_0}-\alpha\,H_{{K}}&-j_{{{\it ex}}}-H_{{K}
}+\alpha\,{\it I_0}\\ \noalign{\medskip}-j_{{{\it ex}}}&\alpha\,j_{{{
\it ex}}}&j_{{{\it ex}}}+H_{{K}}-\alpha\,{\it I_0}&-\alpha\,j_{{{\it ex
}}}-{\it I_0}-\alpha\,H_{{K}}\end {array} \right] \]
\end{widetext}
whose critical eigenvalues can be reduced to $\Re\left({\lambda_{1,2}}\right)=-\alpha \ H_K  + I_0 $. $\Re\left({\lambda_{3,4}}\right)$ become positive for  larger $I_0$ values, hence do not affect stability.   Solving for $I_0$  that makes the real part positive, we obtain the critical switching current for 1 and 2  as $ | I^c_{Si} | = q \gamma \alpha H_K N_i $. Adding both currents to these layers gives Eq.~(\ref{eq:singlefm}) for the total current since $N_i = (M_s V)_i  \ \mu_B$  and $\gamma = 2\mu_B/\hbar$. The critical current for switching from $+1$ is obtained similarly, giving the same absolute threshold. 

\textbf{Synthetic AFM [Symmetric Currents] (Eq.~(\ref{eq:AFMc}))} \\
Let  $V_2/V_1 = N_2/ N_1= \beta < 1$ and $\vec{I}_{S1}/ (\gamma q N_1)=\vec{I}_{S2}/ ( \gamma q N_2)=I_0 \hat z$ with the initial condition $m_{z1}=-1 \ m_{z2}=+1$, and $\vec{H}_1= H_{K} m_{z1} \  \hat z - \beta j_{ex}  \hat m_2$, $\vec{H}_2= H_{K} m_{z2} \  \hat z - j_{ex}  \hat m_1$,and $-j_{ex}<0$ for AFM coupling.  The reduced Jacobian then becomes:
\begin{widetext}
\[J_r= \left[\begin {array}{cccc} -\beta\,j_{{{\it ex}}}\alpha+{\it I_0}-
\alpha\,H_{{K}}&\beta\,j_{{{\it ex}}}+H_{{K}}+\alpha\,{\it I_0}&-\beta
\,j_{{{\it ex}}}\alpha&\beta\,j_{{{\it ex}}}\\ \noalign{\medskip}-
\beta\,j_{{{\it ex}}}-H_{{K}}-\alpha\,{\it I_0}&-\beta\,j_{{{\it ex}}}
\alpha+{\it I_0}-\alpha\,H_{{K}}&-\beta\,j_{{{\it ex}}}&-\beta\,j_{{{
\it ex}}}\alpha\\ \noalign{\medskip}-\alpha\,j_{{{\it ex}}}&-j_{{{\it 
ex}}}&-\alpha\,j_{{{\it ex}}}-{\it I_0}-\alpha\,H_{{K}}&-j_{{{\it ex}}}
-H_{{K}}+\alpha\,{\it I_0}\\ \noalign{\medskip}j_{{{\it ex}}}&-\alpha\,
j_{{{\it ex}}}&j_{{{\it ex}}}+H_{{K}}-\alpha\,{\it I_0}&-\alpha\,j_{{{
\it ex}}}-{\it I_0}-\alpha\,H_{{K}}\end {array} \right] \]
\end{widetext}
We solve for $I_0$ after finding the real part of the eigenvalues $\Re\left({\lambda_{1,2,3,4}}\right)$ that are all degenerate and  picking the smaller root of the two solutions:
\begin{widetext}
\begin{equation} I_0 =\frac{\alpha}{2}\left(\,\beta\,j_{{{\it ex}}}-\,j_{{{\it ex}}}+\,\sqrt {{\beta}^{2}{j_{{{\it ex}}}}^{2}-2
\,{j_{{{\it ex}}}}^{2}\beta+{j_{{{\it 
ex}}}}^{2}+4\,\beta\,j_{{{\it ex}}}H_{{K}}+4\,j_{{{\it ex}}}H_{{K}}+4
\,{H_{{K}}}^{2}} \right) \nonumber
\end{equation}
\label{eq:sup2}
\end{widetext}
Substituting $I_{Si}=I_0/(\gamma q N_i)$, defining $I_{ex}$ and $I_{c}$ as 
\[\begin{array}{l}
{I_{ex}} = \displaystyle\frac{{2q}}{\hbar }\alpha \left( {{J_{ex}}S} \right)\\
\\
{I_c} =  \displaystyle\frac{{2q}}{\hbar }\alpha \left( {2{\rm{ }}{K_F}{V_F}} \right)
\end{array}\]
and factoring the inside of the square root we obtain Eq.~(\ref{eq:AFMc}), which is:
\begin{equation}
{I^c_{S2}} = \frac{{{(\beta  - 1}){I_{ex}}}}{2} \hspace{-0pt}+\hspace{-0pt} \sqrt {\hspace{-0pt}I_c^{}\left( {{I_c} + {I_{ex}}\left( {\beta  \hspace{-0pt}+ \hspace{-0pt}1} \right)} \right)\hspace{-0pt} +\hspace{-0pt} \frac{{I_{ex}^{2}}}{4}{{\left( {\beta \hspace{-0pt} -\hspace{-0pt} 1} \right)}^2}} 
\end{equation}

 In the limit  $\beta=N_1/N_2=1$ we obtain:
\[
I^c_{S0} = \sqrt{I_c^2+ 2 I_c I_{ex}}
\] 
as in Eq.~(\ref{eq:afm}).

\textbf{Synthetic AFM [Anti-Symmetric Currents] (Eq.~(\ref{eq:afm_asym}))} \\
In this section we derive Eq.~(\ref{eq:afm_asym}) in the equal volume ($\beta=1$) limit.  The initial condition $m_{z1}=+1 \ m_{z2}=-1$ but now   $-\vec{I}_{S1}/ (\gamma q N)=\vec{I}_{S2}/ ( \gamma q N)=I_0 \hat z$. The reduced Jacobian becomes:
\begin{widetext}
\[J_r = \left[ \begin {array}{cccc} -j_{{{\it ex}}}{\it \alpha}+{\it I_0}-{\it 
\alpha}\,H_{{K}}&j_{{{\it ex}}}+H_{{K}}+{\it \alpha}\,{\it I_0}&-j_{{{
\it ex}}}{\it \alpha}&j_{{{\it ex}}}\\ \noalign{\medskip}-j_{{{\it ex}}
}-H_{{K}}-{\it \alpha}\,{\it I_0}&-j_{{{\it ex}}}{\it \alpha}+{\it I_0}-{
\it \alpha}\,H_{{K}}&-j_{{{\it ex}}}&-j_{{{\it ex}}}{\it \alpha}
\\ \noalign{\medskip}-j_{{{\it ex}}}{\it \alpha}&-j_{{{\it ex}}}&-j_{{{
\it ex}}}{\it \alpha}+{\it I_0}-{\it \alpha}\,H_{{K}}&-j_{{{\it ex}}}-H_{
{K}}-{\it \alpha}\,{\it I_0}\\ \noalign{\medskip}j_{{{\it ex}}}&-j_{{{
\it ex}}}{\it \alpha}&j_{{{\it ex}}}+H_{{K}}+{\it \alpha}\,{\it I_0}&-j_{
{{\it ex}}}{\it \alpha}+{\it I_0}-{\it \alpha}\,H_{{K}}\end {array}
 \right]\]
\end{widetext}
The real parts of all $\Re\left({\lambda_{1,2,3,4}}\right)$ are degenerate and the critical $I_0$ becomes $ = \alpha H_K + j_{ex} \alpha$, making $I^c_{S0}=(I_{ex}+I_c)$ as in Eq.~(\ref{eq:afm_asym}).

\section{Conservation of Angular Momentum}
In this section, we show the conservation of angular momentum for a single magnet during switching by a spin-current paying special attention to near overdrive and high overdrive regimes. 

Starting from Eq.~(\ref{eq:sup})  for a single, monodomain magnetic layer with perpendicular magnetic anisotropy (PMA) having an easy axis pointing along the  the z-direction, we focus on the z-component of magnetization which reduces to \cite{datta2012new} :
\begin{equation}
qN_s \frac{dm_z}{dt}  = \frac{1-m_z^2}{1+\alpha^2} \left( I_s + I_{sc} \  m_z \right)
\label{eq:supeq}
\end{equation}
where $I_{sc}= q N_s \alpha \gamma H_K $ and $N_s= (M_s \rm V)/\mu_B$, $\mu_B$ being the Bohr magneton.  This nonlinear ODE does not have a direct, closed form solution, although implicit solutions exist. 

We are interested in the total change of angular momentum in the z-direction during switching from an initial state of $-1<m_i<0$ to a final state of $m_f=1$.
Therefore integrating both sides of Eq.~(\ref{eq:supeq}):
\begin{equation}
qN_s \underbrace{\int_{m_i=m_0}^{m_f=+1} d m_z}_{ 1  -m_0} = \int_{t=0}^{t=\infty}\frac{1-m_z^2}{1+\alpha^2} \left(\underbrace{I_s}_{\rm stt} + \underbrace{I_{sc} \  m_z}_{\rm damping} \right) 
\end{equation}
where we identified the spin-transfer-torque and damping contributions to the net angular momentum transfer. In the limit that spin-transfer-torque dominates ($I_s \gg  I_{sc} \ |m_z |$, noting $|m_z| \le 1$ at all times),  the damping term can be ignored as noted by \cite{behin2011switching}. In this case  most of the angular momentum $q N_s (1-m_0)\approx 2qN_s$ is provided by a spin-current. In general the damping term cannot be ignored and in the absence of a  spin-torque input, the damping provides all of the angular momentum, for instance in the case of a magnet that relaxes from its hard axis to its easy axis.

In addition,  numerical simulations show that in the low overdrive regime ($I_s \gtrsim I_{sc}$), the damping can hinder the spin-torque switching while an initial deviation is building up, causing spin-torque current to deposit an amount of angular momentum that can be significantly greater than $2qN_s$. 

\bibliography{Exchange2015.bib}

\clearpage

\clearpage

\clearpage

\clearpage

\clearpage

\end{document}